\documentclass[aip,amsmath,amssymb,reprint]{revtex4-2}

\usepackage{graphicx}
\usepackage{dcolumn}
\usepackage{bm}
\usepackage[utf8]{inputenc}
\usepackage[T1]{fontenc}
\usepackage{mathptmx}
\usepackage{booktabs}
\usepackage{color}
\usepackage{multirow}
\usepackage{physics}
\usepackage{pgfplots}
\usepackage[version=4]{mhchem}
\usepackage{parskip}
\usepackage[caption=false]{subfig}
\newcommand{\angstrom}{\text{\normalfont\AA}}

\begin{document}

\title{Outstanding Improvement in Removing the Delocalization Error by Global Natural Orbital Functional}

\author{Juan Felipe Huan Lew-Yee}
\affiliation{Departamento de F\'isica y Qu\'imica Te\'orica, Facultad de Qu\'imica, Universidad Nacional Aut\'onoma de M\'exico, M\'exico City, C.P. 04510, M\'exico}

\author{Mario Piris}
\email{mario.piris@ehu.eus}
\affiliation{Donostia International Physics Center (DIPC), 20018 Donostia, Euskadi, Spain; Euskal Herriko Unibertsitatea (UPV/EHU), PK 1072, 20080 Donostia, Euskadi, Spain; and Basque Foundation for Science (IKERBASQUE), 48009 Bilbao, Euskadi, Spain}

\author{Jorge M. del Campo}
\email{jmdelc@unam.mx}
\affiliation{Departamento de F\'isica y Qu\'imica Te\'orica, Facultad de Qu\'imica, Universidad Nacional Aut\'onoma de M\'exico, M\'exico City, C.P. 04510, M\'exico}

\date{\today}

\begin{abstract}

This work assesses the performance of the recently proposed global natural orbital functional (GNOF) against the charge delocalization error. GNOF provides a good balance between static and dynamic electronic correlation leading to accurate total energies while preserving spin, even for systems with a highly multi-configurational character. Several analyses were applied to the functional, namely i) how the charge is distributed in super-systems of two fragments, ii) the stability of ionization potentials while increasing the system size, and iii) potential energy curves of a neutral and charged diatomic system. GNOF was found to practically eliminate the charge delocalization error in many of the studied systems or greatly improves the results obtained previously with PNOF7.

\end{abstract}

\maketitle

\section{Introduction}

The delocalization error is one of the biggest pending challenges of current electronic structure methods.\cite{Bryenton2022-zd,Cohen2008-so} It appears in many density functional approximations (DFAs),\citep{Mori-Sanchez2006-bz,Hait2018-jg} as well as wavefunction-based methods, such as unrestricted Hartree-Fock and its second-order perturbative corrections.\citep{Cohen2009-wj} It has been studied from many perspectives considering that the energy of a system with a fractional number of electrons should be linearly related to the energy of the closest systems with integer electron numbers. Consequently, the lack of discontinuity in the energy derivative at an integer number of particles correlates with the delocalization error.\citep{Perdew1982-qk} The delocalization error has been explained as a self-interaction consequence,\citep{Zhang1998-sk} and more recently as a multi-electron self-interaction error.\citep{Mori-Sanchez2006-bz} This problem causes an energetic overstabilization of fractional charges\citep{Cohen2007-pi}, and has important consequences in many chemical situations of interest as band gap predictions\citep{Mori-Sanchez2008-lj}, dissociation curves\citep{Ruzsinszky2006-eu} of neutral and charged molecules,\citep{Ruzsinszky2005-ry} and ionization potentials.\citep{Whittleton2015-xt} Due to its relevance to chemical predictions, many attempts have been made to overcome the problem of the delocalization error in DFAs.\citep{Komsa2016-ba} Some of these ideas include functional tuning,\citep{Whittleton2015-xt, Autschbach2014-nu} the design of explicitly corrected functionals\citep{Cohen2007-pi, Becke2018-tv, Proynov2021-oy}, and the use of machine learning approaches.\citep{Kirkpatrick2021-np, Perdew2021-is} In fact, surpassing this error has become a desirable feature to be satisfied in the development of new methods.\citep{Bao2017-vo}

One-particle reduced density matrix (1RDM) functional theory emerged\citep{Gilbert1975,Levy1979,Valone1980} in the 1970s as an alternative formalism to both density functional and wavefunction based methods. Advances in this area lead to approximate functionals of the 1RDM in its diagonal form, that is, the use of natural orbitals (NOs) and its occupation numbers (ONs), which define a natural orbital functional (NOF).\citep{Goedecker2000,Piris2007} It is more appropriate to speak of a NOF rather than a 1RDM functional when dealing with approximations, since a dependency of the two-particle RDM (2RDM) persists\citep{Donnelly1979} leading to the functional N-representability problem.\citep{Ludena2013,Piris2018d} Comprehensive reviews of approximate NOFs can be found elsewhere.\citep{Pernal2016, Schade2017, Mitxelena2019} Nowadays, the NOF theory has become an active field of research,\citep{Piris2019, Benavides-Riveros2019, Cioslowski2019, Giesbertz2019, Gritsenko2019, Lopez2019, Schilling2019, Quintero-Monsebaiz2019, Mitxelena2020, Mitxelena2020b, Giesbertz2020, Cioslowski2020, Mitxelena2020c, Mercero2021, Quintero-Monsebaiz2021, Rodriguez-Mayorga2021, Schilling2021,  Piris2021b, DiSabatino2022, Mitxelena2022, Liebert2022, Wang2022, Rodriguez-Mayorga2022, Senjean2022} and several advances have been achieved regarding to its efficient implementation.\citep{Lew-Yee2021,Yao2021,Yao2022-sa,Lemke2022} Special emphasis should be made on the open source program DoNOF\citep{Piris2021} that has been made available to the community (github.com/DoNOF) to perform NOF-based calculations.

The design of NOFs has been highly motivated by their ability to recover static correlation through fractional ONs, offering an intermediate cost between multireference methods and common DFAs. In fact, approximate NOFs have demonstrated to be more accurate than their electron density-dependent counterparts and to have better scaling with respect to the number of basis functions than wavefunction-type methods for systems with large amount of strong non-dynamic correlation. Particularly successful in describing static electron correlation have been Piris NOFs (PNOFs) based on electron-pairing,\citep{Piris2018e} namely PNOF5,\citep{Piris2011,Piris2013e} PNOF6,\citep{Piris2014c} and PNOF7.\citep{Piris2017,Mitxelena2018a} These NOFs are the only ones that have hitherto been able to achieve the correct number of electrons for the resulting fragments in homolytic dissociations,\citep{Matxain2011,Ruiperez2013} therefore, no delocalization problems have been observed in these processes.

Although some reports have considered the delocalization error in NOFs,\cite{Hellgren2019-tf} this issue has not received much attention as in other electronic structure methods. In a previous report\citep{Lew-Yee2022}, it has been shown that PNOFs can simultaneously deal with static correlation and charge delocalization errors, becoming a promising option for studying charge-related problems. In particular, PNOF5 was shown to prefer localized solutions, whereas PNOF7 can suffer from charge delocalization error, although it far outperforms common DFAs. Two years after PNOF5 was proposed\citep{Piris2011}, it was realized\citep{Pernal2013} that an antisymmetrized product of strongly orthogonal geminals with the expansion coefficients explicitly expressed by the ONs leads to it, which confirms that PNOF5 is strictly N-representable, i.e., the functional can also be derived from a wavefunction that is antisymmetric in N-particles. This exceptional property of PNOF5 is responsible for the absence of delocalization error, as occurs with wavefunctions that do not violate inherent physical symmetries. Nevertheless, we must recall that PNOF5 is equivalent to an independent electron pair model, hence it only takes into account the intrapair electron correlation, whereas PNOF7 also includes essentially non-dynamic interpair correlation which allows it to perform better on challenging strong correlation problems\citep{Mitxelena2017a,Quintero-Monsebaiz2019,Mitxelena2020,Mitxelena2020b} where PNOF5 fails.

Like in the other PNOFs, PNOF7 satisfies several analytic (2,2)-positivity conditions\citep{Mazziotti2012a} as a consequence of imposing them on the reconstructed 2RDM. It is well known that these conditions are necessary but not sufficient for the ensemble N-representability, so there might be situations where PNOF7 violates the N-representability and the delocalization error appears. This small but consistent charge delocalization error shown by PNOF7 was related\citep{Lew-Yee2022} to spurious contributions of static correlation due to the absence of dynamic interpair correlation terms in the functional. We therefore hope that a better balance between dynamic and static electron correlation will improve or even make the delocalization error disappear.

Recently,\citep{Piris2021b} a new NOF has been proposed for electronic systems with any spin value regardless of the external potential, i.e. a global NOF (GNOF), to precisely achieve a better balance of static and dynamic electronic correlation even for those systems with multi-configurational character, preserving total spin. The agreement obtained\citep{Mitxelena2022} by GNOF with accurate wavefunction-based methods is good for relative energies and for absolute energies, a fact that points out that good results come out for good reasons. Therefore, it invites us to test its performance in the delocalization error, and the best correlation balance is expected to provide improved results. The objective of this work is to show that the dynamic correlation provided by GNOF allows to greatly improve the performance in the delocalization problem.

The work is organized as follows. First, Section~\ref{sec:theory} presents a brief review of GNOF. This is followed by the computational details of the tests used to characterize the charge delocalization error in Section~\ref{sec:comp-details}. Section~\ref{sec:results} presents an analysis of the performance of GNOF over the charge delocalization error in dimers, ionization potentials of chains and the fractional charge that arises in the dissociation of diatomic molecules. Finally, conclusions are given in Section~\ref{sec:conclusions}.

\section{Theory}\label{sec:theory}

In this section, we briefly describe GNOF, a more detailed description can be found in Ref.~\citep{Piris2021b} The nonrelativistic Hamiltonian under consideration is spin coordinate free; therefore, a state with total spin $S$ is a multiplet, i.e., a mixed quantum state that allows all possible $S_{z}$ values. We consider $\mathrm{N_{I}}$ single electrons which determine the spin $S$ of the system, and the rest of electrons ($\mathrm{N_{II}}=\mathrm{N-N_{I}}$) are spin-paired, so that all spins corresponding to $\mathrm{N_{II}}$ electrons altogether provide a zero spin. In the absence of single electrons ($\mathrm{N_{I}}=0$), the energy reduces to a NOF that describes singlet states, as expected.

We focus on the mixed state of highest multiplicity: $2S+1=\mathrm{N_{I}}+1,\,S=\mathrm{N_{I}}/2$.\citep{Piris2019} For an ensemble of pure states $\left\{ \left|SM_{s}\right\rangle \right\} $, we note that the expected value of $\hat{S}_{z}$ for the whole ensemble is zero. Consequently, the spin-restricted theory can be adopted even if the total spin of the system is not zero. We use a single set of orbitals for $\alpha$ and $\beta$ spins. All the spatial orbitals will then be doubly occupied in the ensemble, so that occupancies for particles with $\alpha$ and $\beta$ spins are equal: $n_{p}^{\alpha}=n_{p}^{\beta}=n_{p}$. 

The orbital space $\Omega$ is divided into two subspaces: $\Omega=\Omega_{\mathrm{I}}\oplus\Omega_{\mathrm{II}}$. $\Omega_{\mathrm{II}}$ is composed of $\mathrm{N_{II}}/2$ mutually disjoint subspaces $\Omega{}_{g}$. Each of which contains one orbital $\left|g\right\rangle $ with $g\leq\mathrm{N_{II}}/2$, and $\mathrm{N}_{g}$ orbitals $\left|p\right\rangle $ with $p>\mathrm{N_{II}}/2$, namely,
\begin{equation} 
\Omega{}_{g}=\left\{ \left|g\right\rangle ,\left|p_{1}\right\rangle ,\left|p_{2}\right\rangle ,...,\left|p_{\mathrm{N}_{g}}\right\rangle \right\} .\label{OmegaG}
\end{equation}
Taking into account the spin, the total occupancy for a given subspace $\Omega{}_{g}$ is 2. In general, $\mathrm{N}_{g}$ can be different for each subspace as long as it describes the electron pair well. For convenience, in this work, we take it equal for all subspaces $\Omega{}_{g}\in\Omega_{\mathrm{II}}$ to the maximum possible value determined by the basis set used in calculations.

Similarly, $\Omega_{\mathrm{I}}$ is composed of $\mathrm{N_{I}}$ mutually disjoint subspaces $\Omega{}_{g}$. In contrast to $\Omega_{\mathrm{II}}$, each subspace $\Omega{}_{g}\in\Omega_{\mathrm{I}}$ contains only one orbital $g$ with $2n_{g}=1$. It is worth noting that each orbital is completely occupied individually, but we do not know whether the electron has $\alpha$ or $\beta$ spin: $n_{g}^{\alpha}=n_{g}^{\beta}=n_{g}=1/2$. 

Reconstruction of 2RDM in terms of ONs leads to GNOF:
\begin{equation}
E=E^{intra}+E_{HF}^{inter}+E_{sta}^{inter}+E_{dyn}^{inter} \label{gnof}
\end{equation}

The intra-pair component is formed by the sum of the energies of the electron pairs with opposite spins and the single-electron energies of the unpaired electrons, namely 
\begin{equation}
E^{intra}=\sum\limits _{g=1}^{\mathrm{N_{II}}/2}E_{g}+{\displaystyle \sum_{g=\mathrm{N_{II}}/2+1}^{\mathrm{N}_{\Omega}}}H_{gg}
\end{equation}
\begin{equation}
E_{g}=\sum\limits _{p\in\Omega_{g}}n_{p}(2H_{pp}+J_{pp}) + \sum\limits _{q,p\in\Omega_{g},p\neq q}\Pi\left(n_{q},n_{p}\right)L_{pq}
\end{equation} 
\begin{equation}
\Pi\left(n_{q},n_{p}\right)=\sqrt{n_{q}n_{p}}\left(\delta_{q\Omega^{a}}\delta_{p\Omega^{a}}-\delta_{qg}-\delta_{pg}\right)
\end{equation}

$H_{pp}$ are the diagonal one-electron matrix elements of the kinetic energy and external potential operators, whereas $J_{pq}=\left\langle pq|pq\right\rangle$ and $L_{pq}=\left\langle pp|qq\right\rangle$ are the Coulomb and exchange-time-inversion integrals, respectively. $\mathrm{\mathrm{N}_{\Omega}=}\mathrm{N_{II}}/2+\mathrm{N_{I}}$ denotes the total number of suspaces in $\Omega$, as $\Omega^{a}$ denotes the subspace composed of orbitals above the level $\mathrm{N}_{\Omega}$ ($p>\mathrm{N}_{\Omega}$). 

The inter-subspace Hartree-Fock (HF) term is 
\begin{equation}
E_{HF}^{inter}=\sum\limits _{p,q=1}^{\mathrm{N}_{B}}\,'\,n_{q}n_{p}\left(2J_{pq}-K_{pq}\right)\label{ehf}
\end{equation}
where $K_{pq}=\left\langle pq|qp\right\rangle $ are the exchange integrals. The prime in the summation indicates that only the inter-subspace terms are taken into account ($p\in\Omega{}_{f},q\in\Omega{}_{g},f\neq g$). $\mathrm{N}_{B}$ represents the number of basis functions considered. The inter-subspace static component is written as 
\begin{equation}
\begin{array}{c}
E_{sta}^{inter}=-\left({\displaystyle \sum_{p=1}^{\mathrm{N}_{\Omega}}\sum_{q=\mathrm{N}_{\Omega}+1}^{\mathrm{N}_{B}}+\sum_{p=\mathrm{N}_{\Omega}+1}^{\mathrm{N}_{B}}\sum_{q=1}^{\mathrm{N}_{\Omega}}}\right.\left.{\displaystyle +\sum_{p,q=\mathrm{N}_{\Omega}+1}^{\mathrm{N}_{B}}}\right)'\\
\\
\Phi_{q}\Phi_{p}L_{pq}-\:\dfrac{1}{2}\left({\displaystyle \sum\limits _{p=1}^{\mathrm{N_{II}}/2}\sum_{q=\mathrm{N_{II}}/2+1}^{\mathrm{N}_{\Omega}}+\sum_{p=\mathrm{N_{II}}/2+1}^{\mathrm{N}_{\Omega}}\sum\limits _{q=1}^{\mathrm{N_{II}}/2}}\right)'\\
\\
\Phi_{q}\Phi_{p}L_{pq}{\displaystyle \:-\:\dfrac{1}{4}\sum_{p,q=\mathrm{N_{II}}/2+1}^{\mathrm{N}_{\Omega}}}K_{pq}
\end{array} \label{esta}
\end{equation}
where $\Phi_{p}=\sqrt{n_{p}h_{p}}$ with the hole $h_{p}=1-n_{p}$. Note that $\Phi_{p}$ has significant values only when the $n_{p}$ differs substantially from 1 and 0. Finally, the inter-subspace dynamic energy can be conveniently expressed as 
\begin{equation}
E_{dyn}^{inter}=\sum\limits _{p,q=1}^{\mathrm{N}_{B}}\,'\,\left[n_{q}^{d}n_{p}^{d}+\;\Pi\left(n_{q}^{d},n_{p}^{d}\right)\right]
\left[1-\delta_{q\Omega_{II}^{b}}\delta_{p\Omega_{II}^{b}}\right]L_{pq}
\label{edyn}
\end{equation}
where $\Omega_{II}^{b}$ denotes the subspace composed of orbitals below the level $\mathrm{N_{II}}/2$, so interactions between orbitals belonging to $\Omega_{II}^{b}$ are excluded from $E_{dyn}^{inter}$. The dynamic part of $n_{p}$ is defined as
\begin{equation}
n_{p}^{d}=n_{p}\cdot e^{-\left(\dfrac{h_{g}}{h_{c}}\right)^{2}},\quad p\in\Omega_{g}\ \label{dyn-on}
\end{equation}
with $h_{c}=0.02\sqrt{2}$. The maximum value of $n_{p}^{d}$ is around 0.012 in accordance with the Pulay\textquoteright s criterion that establishes an occupancy deviation of approximately 0.01 with respect to 1 or 0 for a NO to contribute to the dynamic correlation.

It is worth pointing out that GNOF preserves the total spin of a multiplet: $\expval{\hat{S}^{2}}=S\left(S+1\right)$, and, Eq. (\ref{gnof}) reduces to a PNOF7-like functional when the inter-pair dynamic term ($E_{dyn}^{inter}$) is neglected, and to PNOF5 if the inter-subspace static term ($E_{sta}^{inter})$ is also disregarded.

\section{Computational Details}\label{sec:comp-details}

Several tests related to the charge delocalization error will be applied to the functionals, some related to ionized supersystems of repeated well-separated fragments. To analyze the calculations performed, let us recall the charge localization metric (CLM) proposed in previous work,\citep{Lew-Yee2022} defined as the difference between the most charged fragment and the least charged fragment,
\begin{equation}
    \begin{matrix}
    \text{CLM}
    \end{matrix}
     =     
    \max\begin{Bmatrix}
    \text{Fragment}\\
    \text{Charges}
    \end{Bmatrix}
    -
    \min\begin{Bmatrix}
    \text{Fragment}\\
    \text{Charges}
    \end{Bmatrix}
    \>\>,
    \label{metric}
\end{equation}
where the curly braces indicate the set of all fragment charges in a system. According to this metric, systems with the charge concentrated in a single fragment will present a value of 1. Conversely, systems with the charge fully delocalized will have a value of 0 since all fragments are equally charged. Values of CLM between these two limit cases indicate partial localization of the charge. 

The CLM provides information on how the charge is distributed in a molecule but does not quantify whether the charge delocalization error is present. Because fragments in a given system will be well-separated, there should not be interactions between them. In the case of two fragments, the energy of the a supersystem should be the sum of the energy of the neutral fragment, $E_0$, and the energy of the positive charged fragment $E_{+}$. Consequently, the energy deviation of the supersystem from this expected value, 
\begin{equation}
\varDelta E = E_{system} - (E_0 + E_{+}) \>\>,
\end{equation}
together with the charge distribution, can be used to quantify the charge delocalization error in systems made up of two fragments, with a straightforward extension to more fragments when required. 

On the other hand, the charge delocalization error makes the ionization potentials of such supersystems depend on the number of fragments; therefore, the deviation of the ionization potentials in relation to the number of fragments can also be used as a qualitative indicator of the charge delocalization error.

All NOF calculations were performed in an internal Julia version of the DoNOF code (https://github.com/DoNOF),\citep{Piris2021} using the resolution of the identity implementation.\citep{Lew-Yee2021} Since the purpose of this work is to compare PNOF5, PNOF7 and GNOF at its maximum capacity, the extended pairing approach has been used, that is, $\mathrm{N}_{g}$ is equal to the maximum possible value determined by the basis set used in the calculations, namely a cc-pVDZ/cc-pVDZ-jkfit basis set.\cite{Pritchard2019} In addition, DFAs calculations were performed in ORCA software\cite{Neese2012-iy,Neese2022-ez} and Full-CI calculations were performed on the Psi4 software when required for comparison.\cite{Smith2020-yt,David_Sherrill1999-xy}

\section{Results}\label{sec:results}

\subsection*{Charge delocalization error in dimers}

In a previous report,\citep{Lew-Yee2022} PNOF5 and PNOF7 were shown to outperform common DFAs by exhibiting a much lower charge delocalization error. In fact, PNOF5 was shown to be free of the delocalization error, while PNOF7 exhibited a small but consistent error in very strongly correlated systems. To this end, studies of the relationship between charge distribution and energy stabilization were carried out for chains of well-separated repeated fragments. The 17 molecules with multireference character of the set W4-17-MR\citep{Karton2017-nf} were used as basic units of the chains. In this work, we have adopted a similar approach, that is, we have built a supersystem consisting of two fragments separated by a distance of 10 $\angstrom$ using the same set of W4-17-MR molecules. Our intention is to directly compare the charge distribution and energy stabilization between two fragments obtained by GNOF with the results of PNOF7.

\begin{figure}[htbp]
    \centering
    \includegraphics[width=0.5\textwidth]{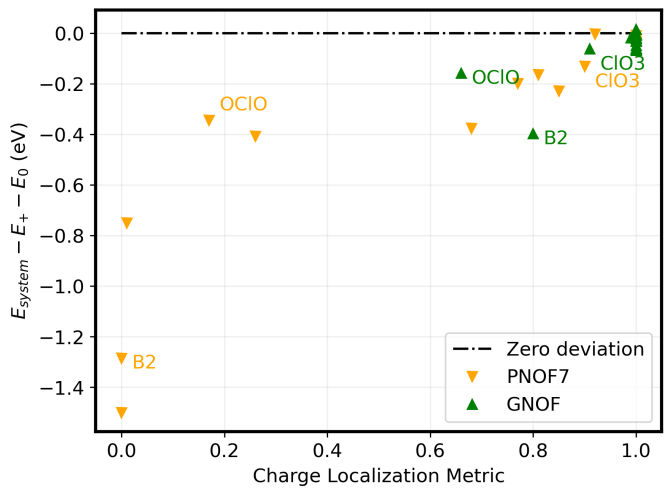}
    \caption{Energy deviation of the supersystems with respect to the energy sum of the neutral and charged fragments, $\varDelta E = E_{system} - (E_0 + E_{+})$, as a function of the charge localization metric.}
    \label{fig:chare-SE}
\end{figure}

The results are presented in Fig.~\ref{fig:chare-SE}, where the vertical axis corresponds to the energy deviation of the supersystem with respect to the energy sum of the isolated charged and neutral fragments, while the horizontal axis corresponds to the CLM. The PNOF7 values are consistent with the previous report.\citep{Lew-Yee2022} Recall that the orange downward triangles on the left (CLM=0) indicate that the charge is equally shared between the fragments, on the right (CLM=1) that the charge is on a single fragment, and those distributed along the horizontal axis indicate that the charge is delocalized between the fragments, increasing the energy deviation with this delocalization. In contrast, the green upward triangles of the GNOF values are located mostly to the right and above, indicating that the charge is located on a single fragment. Regrettably, there are still some systems in which the charge delocalization error has persisted, namely the supersystems \ce{OClO} (CLM = 0.66), \ce{B2} (CLM = 0.80) and \ce{ClO3} (CLM = 0.91). Interestingly, in the case of \ce{OClO}, another solution was also found with the charge located in a single fragment but with higher energy.

\subsection*{Ionization potentials of chains}

A known effect of the charge delocalization error is the deviation of the ionization potentials as the number of fragments in the system increases.\citep{Whittleton2015-xt} Although it has been shown\citep{Lew-Yee2022} that PNOF7 can provide stable results for helium atom chains and other weakly correlated systems, even with a multireference character, its performance deteriorates in cases of extreme static correlation. However, it should be noted that the observed error for PNOF7 is small compared to common DFA errors that can reach several electron volts.\citep{Whittleton2015-xt}

\begin{figure}[htbp]
    \centering
    \includegraphics[width=0.5\textwidth]{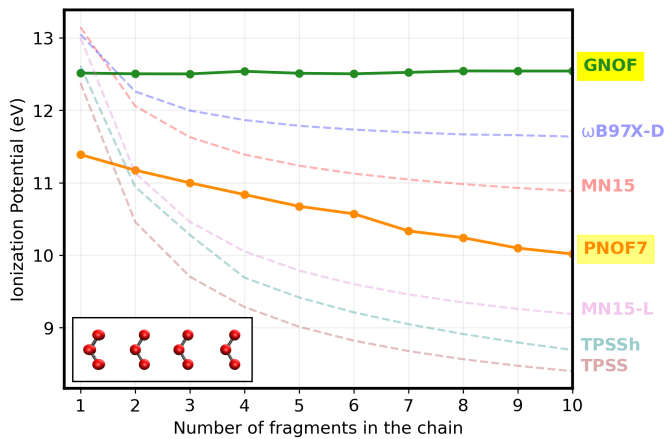}
    \caption{Ionization potentials of chains of repeated and well-separated fragments of \ce{O3}. Each mark correspond to a chain with the number of fragments indicated by the value in the horizontal axis.}
    \label{fig:o3-ionization}
\end{figure}

To illustrate this effect, the ionization potentials of chains formed by repetition of the \ce{O3} molecule were calculated with GNOF, PNOF7, and some common DFAs, as presented in Fig.~\ref{fig:o3-ionization}. For \ce{O3} chains, it was found \cite{Lew-Yee2022} that PNOF7 improves with respect to DFAs in reproducing the ionization potential as the chain size increases, but still fails due to the delocalization error. It can be seen that DFAs present large deviations in ionization potentials; for example, $\omega$B97X-D predicts an ionization potential of 13.0 eV for a single \ce{O3}, but of 11.6 eV for the chain of ten fragments, \ce{(O3)10}, that is, a difference of 1.3 eV. Similar differences can be found in other DFAs as a consequence of the charge delocalization error that affects the ionized system; namely, the aforementioned difference is 2.3 eV for MN15, 3.8 eV for MN15-L, 3.9 eV for TPSSh, and 4.0 eV for TPSS. Note that PNOF7 achieves a 1.3 eV difference between the single molecule and the ten-fragment chain, with an accuracy for a single molecule that is not as good as DFAs, but remains more stable, exceeding the accuracy of some DFAs just beyond the dimer \ce{(O3)2}. By contrast, GNOF predicts stable ionization potentials for all chains. Furthermore, the ionization potential predicted by GNOF is in excellent agreement with the experimental value of 12.5 eV.\citep{Weiss1977}

\subsection*{Potential Energy Curves}

\begin{figure*}[htbp]
    \centering
    \subfloat[\ce{LiH}\label{fig:pec-lih}]{
     \includegraphics[width=0.5\textwidth]{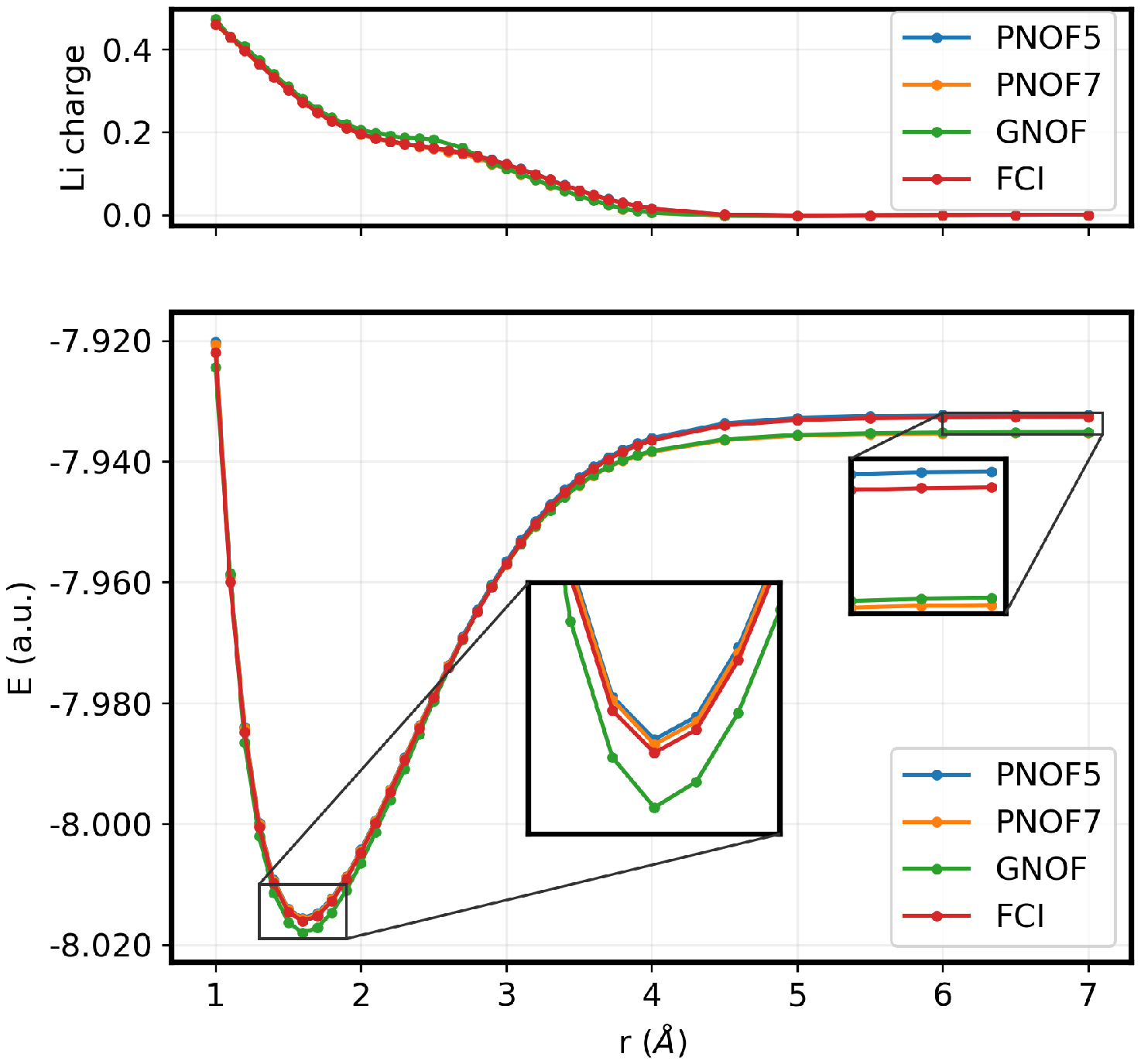}
     }
    \subfloat[\ce{LiH+}\label{fig:pec-lih+}]{
     \includegraphics[width=0.5\textwidth]{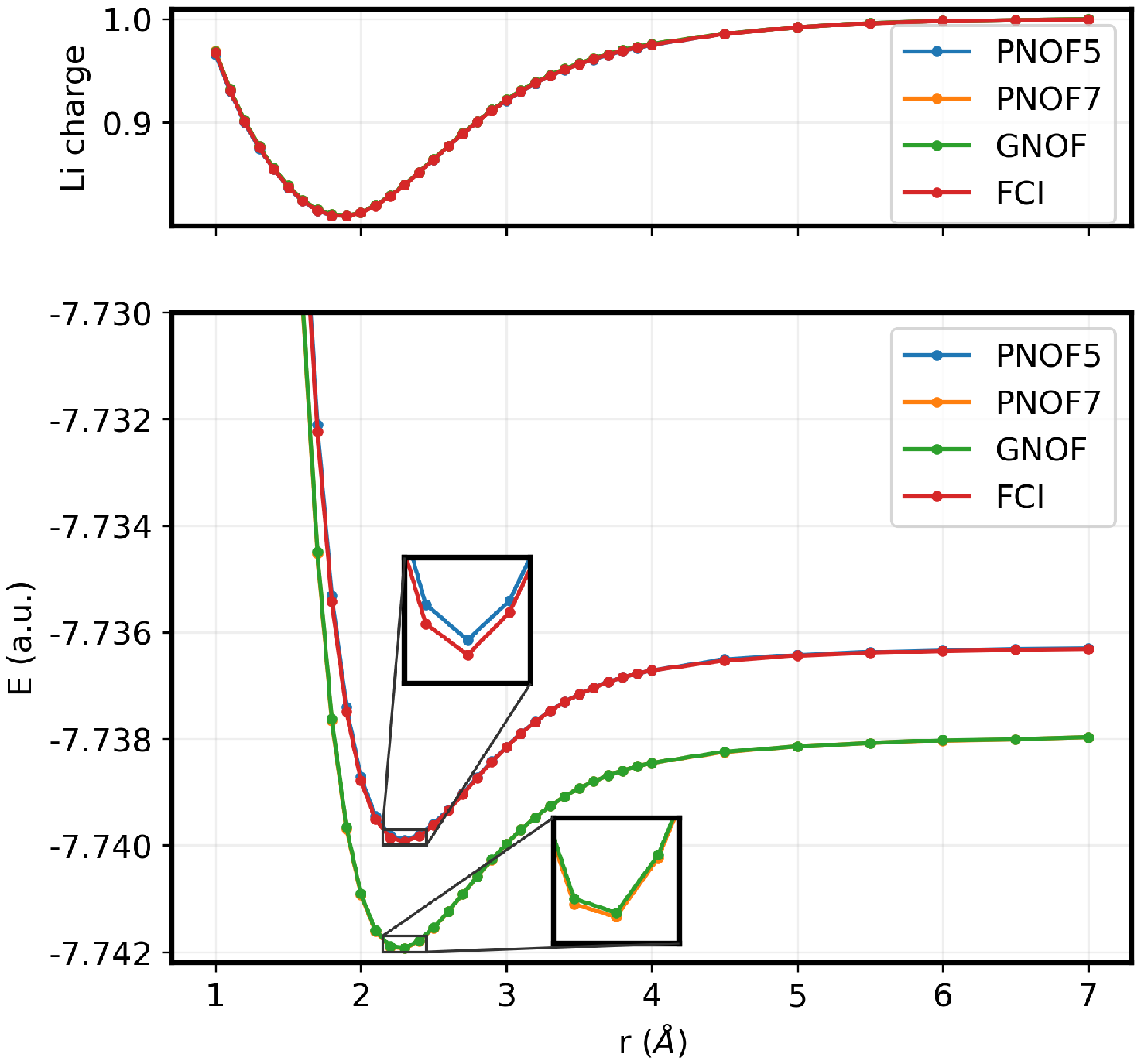}         
     }
    \caption{Dissociation curves of Lithium hydride, (a) neutral (\ce{LiH}) and (b) charged (\ce{LiH^+}). The bottom panels present the potential energy curve, while the top panels present the charge of the lithium atom through the dissociation computed by a L\"owdin population. In Fig.~\ref{fig:pec-lih+}, the PNOF7 (orange) curve is under the curve of GNOF (green), and the PNOF5 (blue) curve is under the curve of FCI (red).}
    \label{fig:lih}
\end{figure*}

Another known problem associated with the charge delocalization error is the fractional charge that arises in the dissociation of diatomic molecules. Take the case of \ce{LiH} as an example, for which it was reported\cite{Hellgren2019-tf} that the charge delocalization error may lead to an incorrect electron distribution in the dissociation limit. The corresponding potential energy curves for neutral and positively charged lithium hydrides are presented in Fig.~\ref{fig:lih} for PNOF5 (blue), PNOF7 (orange) and GNOF (green); also, the FCI curve (re) is presented for comparison.

The neutral \ce{LiH} shown in Fig.~\ref{fig:pec-lih} is formed by four electrons; therefore, the calculation consists of two electron pairs distributed in the double occupied orbitals of the $\Omega_{\mathrm{II}}$ subspace. The bottom panel shows that PNOF5 remains above the FCI curve as a consequence of being exactly N-representable. In contrast, PNOF7 and GNOF may be below the FCI curve due to N-representability violations; nevertheless, GNOF achieves energies close to FCI, with a deviation of 1.6 kcal/mol in the dissociation region (7.0 \AA), and of 1.2 kcal/mol at the bonding region (1.6 \AA). Remarkably, all NOFs dissociate the system into a neutral lithium atom and a neutral hydrogen atom, and the charges along the disociation curve remain almost identical to those predicted by FCI, as can be seen in the top panel of Fig.~\ref{fig:pec-lih}.

The charged \ce{LiH+} is presented in Fig.~\ref{fig:pec-lih+}. This molecule is a three-electron system, then the calculation consists of only one electron pair distributed in the orbitals of the $\Omega_{\mathrm{II}}$ subspace, and one single occupied orbital in the $\Omega_{\mathrm{I}}$ subspace. Again, all NOFs correctly dissociate \ce{LiH+} into a charged lithium atom and a neutral hydrogen atom, with a charge curve that almost perfectly coincides with the FCI curve. Furthermore, PNOF7 and GNOF provide very similar values (the orange curve of PNOF7 is under the green curve of GNOF) below the FCI curve due to N-representability violations, in this case the deviation at the bonding region is 1.3 kcal/mol (2.3 \AA) and 1.0 kcal/mol at the dissociation region (7.0 \AA).

\section{Conclusions}\label{sec:conclusions}

The results provided in this work show that GNOF greatly improves the already promising performance of PNOF7 in the charge delocalization error. In particular, energetic predictions of quantities such as ionization potentials and potential energy curves benefit from an increase in stability. Therefore, the good performance in the charge delocalization error, together with an excellent balance of dynamic and static correlation, makes GNOF a valuable functional for electronic structure calculations of general interest.

\begin{acknowledgments}
Support comes from Ministerio de Economía y Competitividad (Ref. PID2021-126714NB-I00). The authors thank for technical and human support provided by IZO-SGI SGIker of UPV/EHU and European funding (ERDF and ESF). J. F. H. Lew-Yee with CVU Grant No. 867718 acknowledges
CONACyT for the Ph.D. scholarship. J. M. del Campo acknowledges funding from projects Grant Nos. CB-2016-282791, PAPIIT-IN201822, and computing resources from the LANCAD-UNAM-DGTIC-270 project.
\end{acknowledgments}

\textbf{Data Availability}

The data that support the findings of this study are available from the corresponding author upon reasonable request.


%

\end{document}